# **Ignition of a Deuterium Micro-Detonation with a Gigavolt Super Marx Generator**

Friedwardt Winterberg
University of Nevada, Reno
October 2008

**Abstract**

The Centurion – Halite experiment demonstrated the feasibility of igniting a deuterium – tritium micro-explosion with an energy of not more than a few megajoule, and the Mike test the feasibility of a pure deuterium explosion with an energy of more than $10^6$ megajoule. In both cases the ignition energy was supplied by a fission bomb explosive. While an energy of a few megajoule, to be released in the time required of less than $10^{-9}$ sec, can be supplied by lasers and intense particle beams, this is not enough to ignite a pure deuterium explosion. Because the deuterium-tritium reaction depends on the availability of lithium, the non-fission ignition of a pure deuterium fusion reaction would be highly desirable. It is shown that this goal can conceivably be reached with a "super Marx generator", where a large number of "ordinary" Marx generators charge (magnetically insulated) fast high voltage capacitors of a second stage Marx generator, called a "super Marx generator", ultimately reaching gigavolt potentials with an energy output in excess of 100 megajoule. An intense $10^7$ Ampere-GeV proton beam drawn from a "super Marx generator" can ignite a deuterium thermonuclear detonation wave in a compressed deuterium cylinder, where the strong magnetic field of the proton beam entraps the charged fusion reaction products inside the cylinder.

In solving the stand-off problem, the stiffness of a GeV proton beam permits to place the deuterium target at a comparatively large distance from the wall of a cavity confining the deuterium micro-explosion.

## 1. Introduction

Since 1954 I have been actively involved in inertial confinement fusion research, at a time it was still classified in the US. I had independently discovered the basic principles and presented them in 1955 at a meeting of the Max Planck Institute in Goettingen, organized by von Weizsaecker. These principles are the Guderley convergent shock wave and imploding shells solutions to reach high energy densities. The abstracts of the meeting still exist and are kept in the library of the University Stuttgart.

Everyone understands the importance of controlled fusion for the ultimate solution of our energy future, and everyone understands the problem of nuclear waste, which even for deuterium-tritium (DT) fusion still exists, where 80% of the energy goes into neutrons, activating a fusion reactor. Unlike deuterium-tritium fusion which depends on lithium, ordinary water, the raw material for pure deuterium fusion is everywhere abundantly available, and deuterium fusion releases much less energy into neutrons. But unlike DT fusion it requires much higher ignition energies, which I claim can be reached by going to very high voltages, up to a gigavolt.

The goal before us is therefore very clear and can be understood by everyone: It is the attainment in the laboratory of the kind of ultrahigh voltages, as they occur in nature by lightning. The idea to attain inertial confinement by high voltage electric pulse power techniques, instead of lasers, goes back to a paper I had in 1968 published in the Physical Review [1]. In1969 the chairman of the Science Committee of the US House of Representatives Emilio Daddario, declared this idea the most likely to lead to success [2]. Specifically, I had proposed to use a large Marx generator for the ignition of the deuterium-tritium ignition.

Laser fusion will ultimately not work, because for a high gain the intense light flash of a thermonuclear microexplosion is going to destroy the entire optical laser ignition apparatus. The large Livermore laser is intended for weapons simulation. There, a low gain is sufficient.

It is the purpose of this communication to show, that with the proposed super Marx generator one may be able to ignite a pure deuterium thermonuclear micro-explosion.

## 2. Solution in between two extremes

The situation and a solution in between the two extremes is illustrated in Fig. 1. At the lower end one has there the ignition of a DT microexplosion, verified in the Centurion-Halite experiment conducted at the Nevada Test Site, with an ignition energy of a few megajoule. And at the other end, one there has the ignition of a large pure deuterium fusion detonation verified in the Mike test, conducted in the South Pacific, with an ignition energy of more than $10^6$ megajoule.

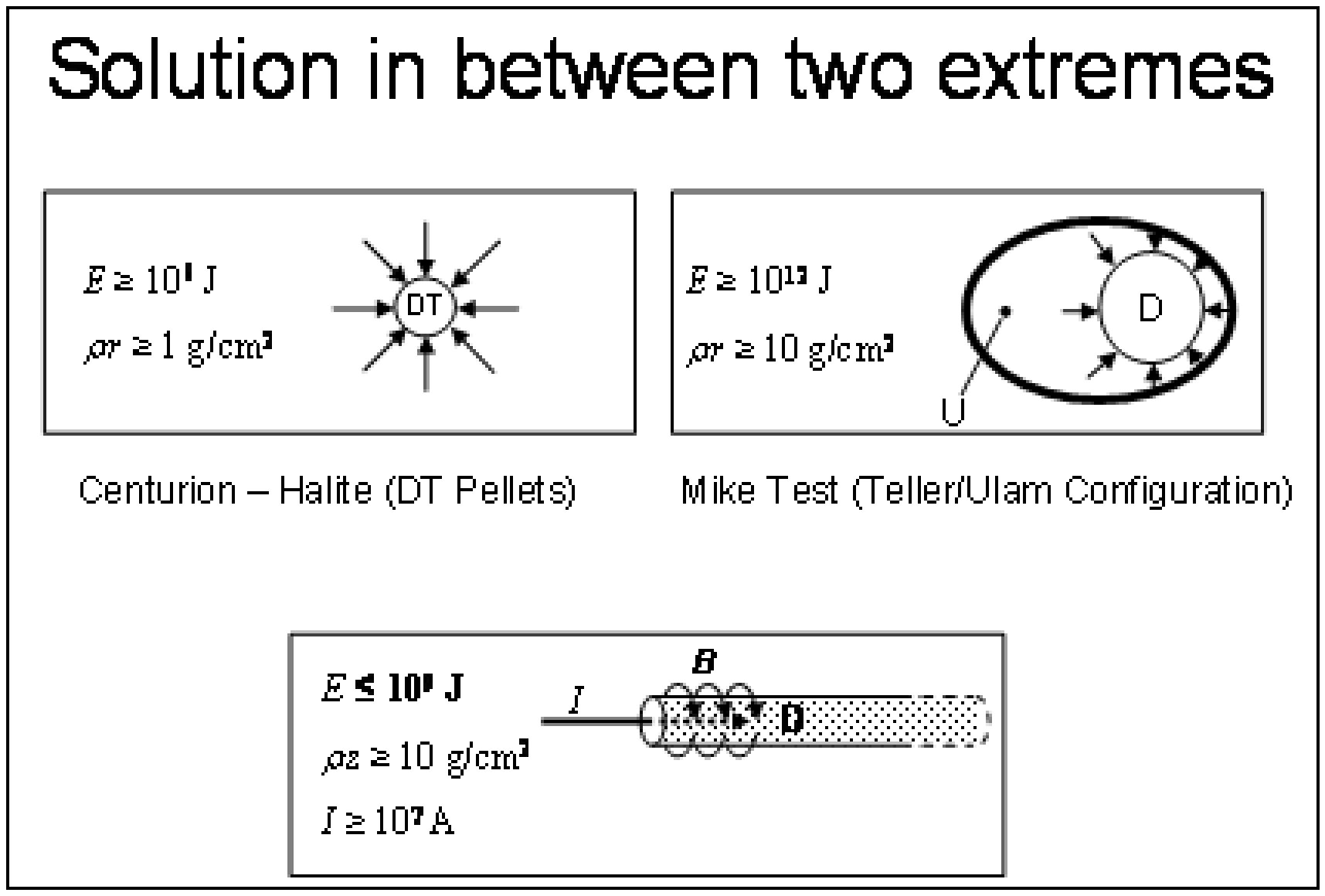

Fig. 1: Ignition of a deuterium target by a GeV-10 MA proton beam.

On a logarithmic scale in between, at an energy of ~ $10^3$ megajoule, there is the hypothetical deuterium configuration for the ignition with the proposed super Marx generator. Assuming an ignition energy of ~ 1 gigajoule, with a gain of 100, the energy output would be 100 gigajoule, corresponding to the energy released by 100 tons of chemical energy. This energy though, is not released into hot gases, as it is in a chemical explosion, but mostly in radiation. And more than 50% is released into charged fusion products, which can be directly converted into electric energy, provided the explosion takes place in a magnetic field filled cavity with a radius of the order 10 meters.

A possible pure deuterium fusion detonation target is shown in Fig. 2, to be ignited by an intense $10^7$ Ampere-GeV proton beam. Part of the energy of the intense proton beam, in entering the target from the left, is scattered by the periphery of a hollow cone, generating a burst of X-rays, ablatively precompressing the deuterium cylinder, placed inside a cylindrical hohlraum. The main portion of the beam energy is focused by the cone onto the deuterium rod, igniting at its end a detonation wave. The beam is stopped over a short distance by the proton-deuterium two stream instability [3], enhanced by a collisionless magnetohydrodynamic shock [4]. At a beam current of ~ $10^7$ Ampere, the charged deuterium fusion reaction products are entrapped within the deuterium cylinder, a necessary condition that a detonation wave propagates down the cylinder.

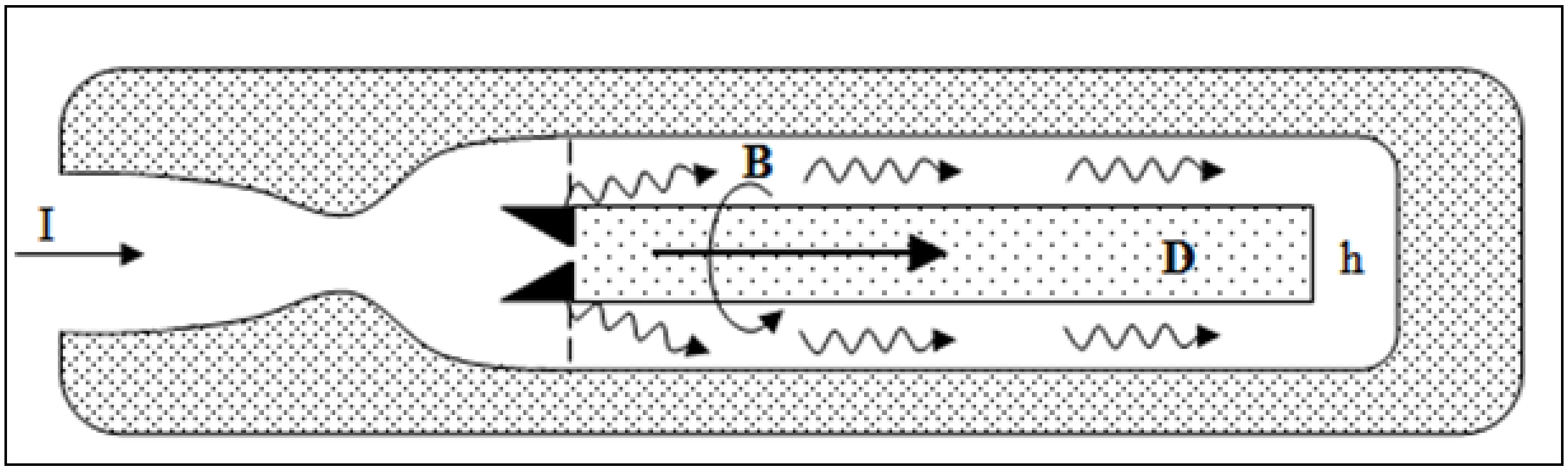

Fig. 2: Possible deuterium micro-detonation target: **I** ion beam, **D** deuterium cylinder, **B** magnetic field, **h** cylindrical hohlraum.

In DT the condition for propagating burn in a spherical target of radius r and density ρ, is given by

$$\rho r \geq 1\text{g/cm}^2 \quad (1)$$

whereas for the D-D reaction the condition is

$$\rho r \geq 10\ \text{g/cm}^2 \quad (2)$$

With the optimal ignition temperature of D-D reaction about 10 times larger than for the DT reaction, and assuming the same density for the compressed deuterium as for deuterium-tritium, it follows that the energy for ignition of a deuterium sphere with a radius 10 times larger than for a deuterium-tritium sphere is $10^4$ larger, that is from a few megajoule for a DT sphere to about $10^4$ megajoule.[1] Obviously, no laser or particle beam can easily reach these kinds of energies. The situation is changed in a fundamental way for a thin deuterium rod of length z, ignited by an intense ion beam with a current of ~ $10^7$ Ampere entrapping the charged fusion reaction products. There the condition (2) is replaced by

[1] With a convergent shock wave ignition in the center of the compressed deuterium sphere this energy is less, but even then still much more than a few megajoule.

$$\rho z \geq 10 \text{ g/cm}^2 \quad (3)$$

Even if the density is less than ~ 100 g/cm$^3$, corresponding to thousandfold compressed liquid hydrogen, the smaller target density can be made up easily by a sufficiently long deuterium cylinder.

The stopping range of the protons by the two stream instability alone is given by

$$\lambda \cong \frac{1.4c}{\varepsilon^{1/3}\omega_i} \quad (4)$$

where $c$ is here is the velocity of light, and $\omega_i$ the proton ion plasma frequency, furthermore $\varepsilon = n_b/n$, where $n_b$ is the proton number density in the proton beam, and n the deuterium target number density. If the cross section of the beam is 0.1 cm$^2$, one obtains for a $10^7$ Ampere beam that $n_b = 2\times10^{16}$ cm$^{-3}$. For a 100 fold compressed deuterium rod one has $n = 5\times10^{24}$ cm$^{-3}$ with $\omega_i = 2\times10^{15}$ s$^{-1}$. One there finds that $\varepsilon = 4\times10^{-9}$ and $\lambda \cong 1.2\times10^{-2}$ cm. This short length, together with the formation of a collision-less magnetohydrogynamic shock, ensures the dissipation of the beam energy into a small volume at the end of the deuterium rod. At a deuterium number density $n = 5\times10^{24}$ cm$^{-3}$, one has $\rho = 17$ g/cm$^{-3}$, and to have $\rho z \geq 10$ g/cm$^{-3}$, thus requires that $z \geq 0.6$ cm. With $\lambda < z$, the condition for the ignition of a thermonuclear detonation wave is satisfied.
With T ≈ $10^9$ K, the ignition energy is given by

$$E_{ign} \sim 3nkT\pi r^2 z \quad (5)$$

For 100 fold compressed deuterium, one has $\pi r^2 = 10^{-2}$ cm$^2$, when initially it was $\pi r^2 = 10^{-1}$ cm$^2$. With $\pi r^2 = 10^{-2}$ cm$^{-2}$, $z$ = 0.6 cm, one finds that $E_{ign} \leq 10^{16}$ erg or ≤ 1 gigajoule. This energy is provided by the $10^7$ Ampere - Gigavolt proton beam lasting $10^{-7}$ seconds. The time is short enough to assure the cold compression of deuterium to high densities. For a $10^3$ fold compression, found feasible in laser fusion experiments, the ignition energy is ten times less.

**3. The Importance of High Voltages for Inertial Confinement Fusion**

The reaching out for high voltages in the quest for the ignition of thermonuclear micro-explosions by inertial confinement can be explained as follows:

1. The energy $e$ [erg] stored in a capacitor $C$ [cm] charged to the voltage $V$ [esu] is equal to

$$e = (1/2)\, CV^2 \quad (6)$$

with an energy density

$$\varepsilon \sim e/C^3 \sim V^2/C^2 \tag{7}$$

The energy $e$ is discharged in the time $\tau$ [sec] ($c$ velocity of light)

$$\tau \sim C/c \tag{8}$$

with the power $P$ [erg/s]

$$P \sim e/\tau \sim cV^2 \tag{9}$$

This shows that for a given dimension of the capacitor measured in its length, and hence volume, the energy stored and power released goes in proportion to the square of the voltage.

2. If the energy stored in the capacitor is released into a charged particle beam with the particles moving at the velocity $v$, the current should be below the critical Alfvén limit:

$$I = \beta\gamma I_A \tag{10}$$

where $\beta = v/c$, $v$ particle velocity, $\gamma = (1 - v^2/c^2)^{-1/2}$ the Lorentz boost factor, and $I_A = mc^3/e$. For electrons $I_A = 17$ kA, but for protons it is 31 MA. If $I << \beta\gamma I_A$, one can view the beam as made up of charged particles accompanied carrying along an electromagnetic field, while for $I >> \beta\gamma I_A$ it is better viewed as an electromagnetic pulse carrying along with it some particles. For $I >> \beta\gamma I_A$, the beam can propagate in a space-charge and current-neutralizing plasma, but only if $I \leq \beta\gamma I_A$ can the beam be easily focused onto a small area, needed to reach a high power flux density. If a power of ~ $10^{15}$ Watt shall be reached with a relativistic electron beam produced by a $10^7$ Volt Marx generator, the beam current would have to be $10^8$ Ampere. For 10MeV electrons one has $\gamma \cong 20$ and $\beta\gamma I_A \sim 3\times10^5$ Ampere, hence $I >> \beta\gamma I_A$. But if the potential is $10^9$ Volt, a proton beam accelerated to this voltage and with a current of $I = 10^7$ A is below the Alfvén current limit for protons, and it would have the power of $10^{16}$ Watt, sufficiently large to ignite a deuterium thermonuclear reaction.

## 4. Super Marx Generator

Would it be not for electric breakdown, one could with a Marx generator reach in principle arbitrarily large voltages. According to Paschen's law, the breakdown voltage in gas between two plane parallel conductors is only a function of the product $pd$, where $p$ is the gas pressure and $d$ the distance between the conductors. For dry air at a pressure of 1 atmosphere the breakdown voltage is $3\times10^4$ V/cm, such that for a pressure of 100 atmospheres the breakdown voltage would be $3\times10^8$ V/cm. For a meter size distance between the conductors this implies a potential difference of the order $10^9$ Volt. But as in lightning, breakdown occurs at much lower voltages by the formation of the “stepped leader”. The formation of a stepped leader though requires some time. Therefore, if the buildup of the high voltage is fast enough, breakdown by a stepped leader can be prevented. In a Marx generator the buildup of the voltage is not fast enough to reach a gigavolt. It is the idea of the super Marx generator how this might be achieved.

To obtain a short discharge time with a single Marx generator, the Marx generator charges up a fast discharge capacitor, discharging its load in a short time. This suggests using a bank of such fast discharge capacitors as the elements of a Marx generator, each one of them charged up by one Marx generator to a high voltage. One may call such a two-stage Marx generator a super Marx generator. If $N$ fast capacitors are charged up by $N$ Marx generators in parallel to the voltage $V$, the closing of the spark gap switches in the super Marx generator adds up their voltages to the voltages $NV$. In the super Marx generator, the Marx generators also serve as the resistors in the original Marx circuit. It is also advantageous to disconnect the Marx generators from the super Marx after its charge- up is completed. Fig. 3 shows the circuit of an ordinary Marx generator, in comparison to a super Marx shown in Fig. 4.

It is known, and used in electric power interrupters, that a high pressure gas flow can blow out a high power electric arc. Vice verse, it can be expected that a rapid gas flow can prevent breakdown [5]. Therefore, just prior to the moment the super Marx generator is fired to one may place the super-Marx generator in a breakdown preventing gas brought into fast motion. An alternative is magnetic insulation in ultrahigh vacuum, by magnetically levitating, and by placing the capacitors of the super Marx in a strong axial magnetic field, insulating the super Marx against radial breakdown. For an axial magnetic field of $B = 2\times10^4 \, Gauss$, the magnetic insulation condition B[Gauss] $\geq$ E[esu] = 300E[Volt/cm], implies magnetic insulation up to $9\times10^7$ Volt/cm. in addition, the current pulse, generated by the closing of the spark gap switches of the super Marx shown in Fig. 4, sets up an axial magnetic field. This too enhances magnetic insulation against radial breakdown. If the first few spark gap switches from the left, are triggered by pulsed lasers, the rise of the voltage at the remaining spark gap switches closes them in an avalanche moving fron the left to the right, leading to a chain of axial current plses moves from the left to the right. The magnitude of the azimuthal magnetic field by these current pulses can be estimated as follows: The energy e = (1/2) $CV^2$, in each of the capacitors charged upto a voltage of ~ $10^7$ Volt is assumed to be e = $10^7$ Joule, which implies that $C = 2\times10^{-7}$ Farad. The charge of each capacitor is Q = CV = 2 Coulomb. With the capacitors consisting of hollow metallic cylinders of length l [cm], the discharge time is of the order $\tau \geq l/c$, (c velocity of light). Assuming that $l \cong 10^3\, cm$, one has $\tau \sim 10^{-7}$ seconds, implying a current, $I = Q/\tau \sim 2\times10^7$ Ampere. At a radius of the cylinder about $R = 3\times10^2$ cm, the magnetic field at this radius is B = 0.2I/R ~ 104 Gauss, sufficiently strong to insulate the (in the vacuum levitated) high voltage capacitors up to $3\times10^6$ V/cm.

The high voltage end of the super Marx has to charge up a magnetically insulated Blumlein transmission line, delivering the GeV proton beam to the deuterium target. The magnetic insulation there can be made by making the Blumlein from co-axial superconducting toruses, as explained in my 1968 Physical Review paper [1].

By connecting the high voltage terminal of the super Marx generator to a Blumlein transmission line, a very high voltage pulse with a fast rise time can be generated. At the envisioned very high voltages one can make a controlled breakdown in a gas, or liquid, generating an ion beam below the Alfvén limit. At these high voltages ion beams are favored over electron beams, because electron beams are there above the Alfvén limit. To assure that all the ions have the same charge to mass ratio, the gas or liquid must be hydrogen or deuterium, otherwise the beam will spread out axially, losing its maximum power.

Instead of making the breakdown in hydrogen gas, one may let the breakdown happen along a thin liquid hydrogen jet, establishing a bridge between the high voltage terminal of the Blumlein transmission line and the thermonuclear target.

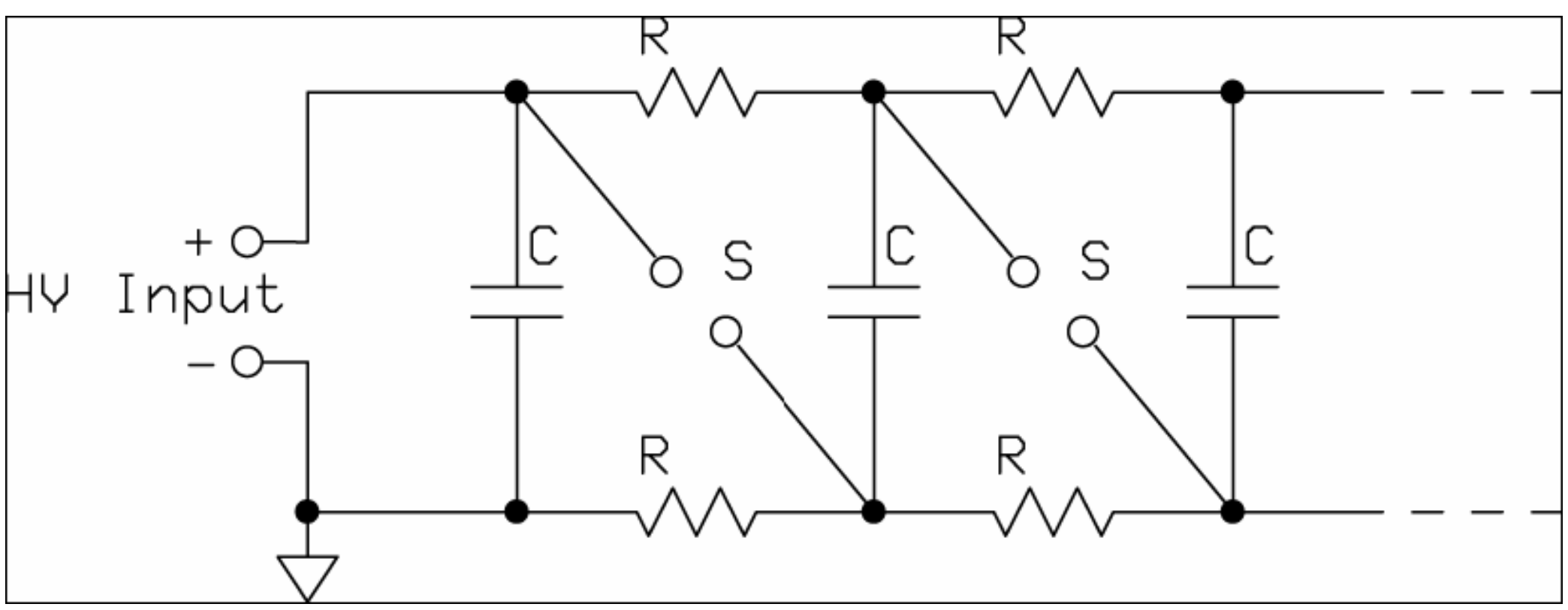

Fig. 3: In an "ordinary" Marx generator n capacitors *C* charged up to the voltage *v*, and are over spark gaps switched into series, adding up their voltages to the voltage *V* = *nv*.

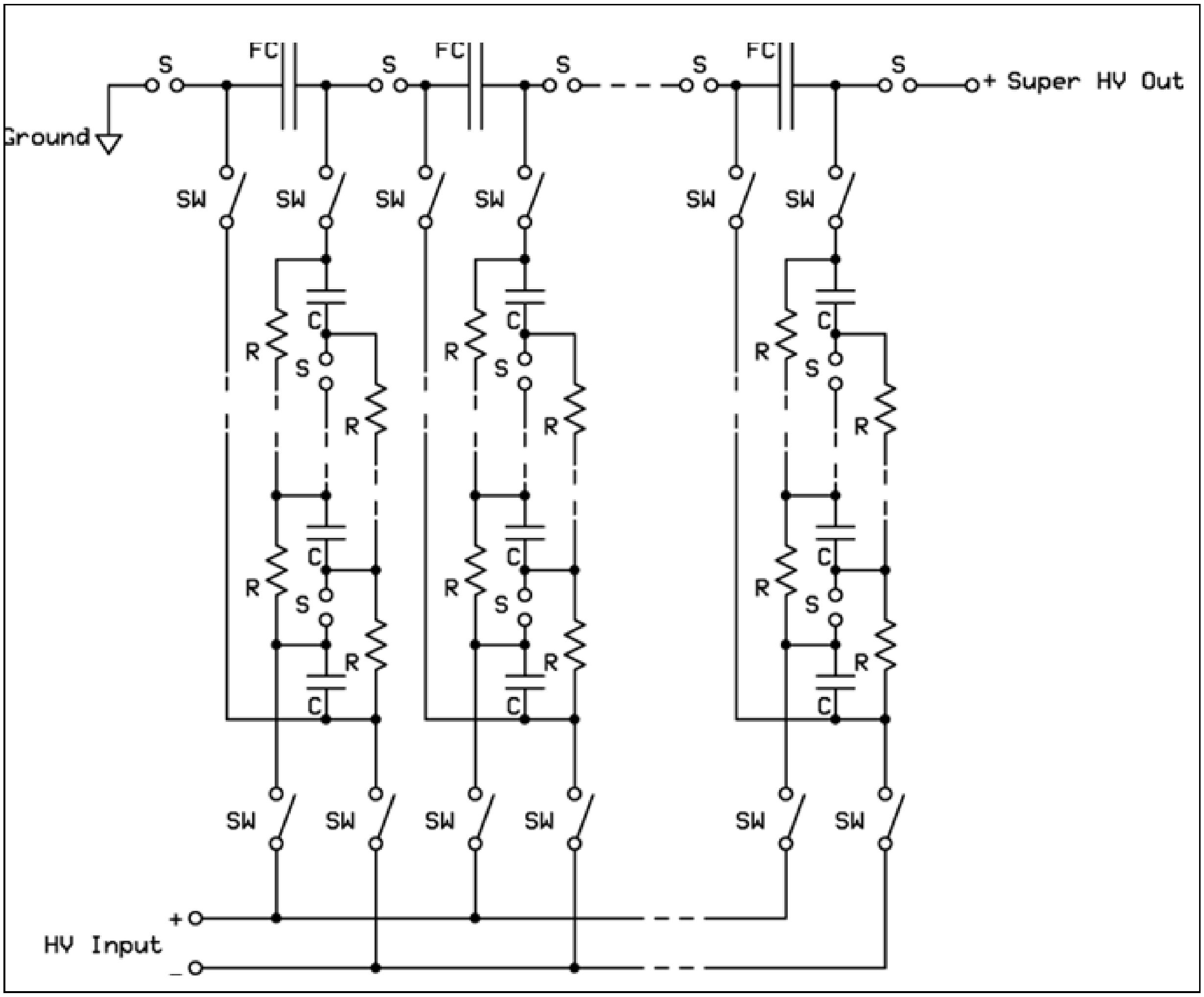

Fig. 4: In a super Marx generator, *N* Marx generators charge up *N* fast capacitors FC to the voltage *V*, which switched into series add up their voltages to the voltage *NV*.

Fig. 5-7 are artistic conceptions of a super Marx generator, and of the chamber where the confined deuterium thermonuclear explosion takes place.

The proposed super Marx generator can reach what nature can do in lightning. The high voltage in natural lightning is released over a distance about 1 km, and the same is true for the super Marx generator.

**Conclusion**

While the ignition without fission of a DT thermonuclear micro-explosion has not yet been achieved, the ignition by a powerful laser beam seems possible in principle. But it is unlikely it will lead to a practical inertial confinement nuclear fusion reactor, because of the intense photon burst of a high gain micro-explosion (required for a inertial confinement fusion reactor), is destroying the laser. Ignition is also likely possible with intense GeV heavy ion beams, but there the stopping of the beam in the target is a problem. In either case, 80% of the energy released goes into the 14 MeV neutrons of the DT reaction. For this reason, the future of DT inertial confinement fusion is likely a hybrid fusion-fission reactor, with fusion providing the neutrons and fission the heat. It favors inertial fusion by encapsulating the DT "pellet" in a U238 or Th232 shell. There the gain not only can be increased through the fission reactions inside the shell, amplifying the implosion of the shell, but it avoids the meltdown problem, which still exists if one surrounds the DT micro-explosion reactor with a subcritical reactor, in particular a subcritical natural uranium light water reactor.

Putting the often cited, neutron-free $HB^{11}$ reaction aside, which under realistically attainable pressures cannot be ignited, leaves us with the reaction of pure deuterium. It not only can be ignited, but permits propagating burn by a detonation wave with the inclusion of T and $He^3$ reaction products [6]. Since the raw material is there ordinary water, everywhere abundantly available, the real challenge for nuclear fusion is in the ignition of pure deuterium.

As it is with rocket technology, what cannot easily achieved in one stage, applies to fusion as well. One could in principle ignite a deuterium fusion reaction with the help of a small DT micro-explosion, but there the targets may become quite complex and thus expensive. Schemes of this kind might still be of interest for pulsed fusion micro-explosion driven space craft, but much less likely for an economically competitive power plant. It is for this reason that the ultimate goal of controlled fusion research should be directed towards the development of by orders of magnitude more powerful drivers, with the ultimate goal to have drivers powerful enough to ignite pure deuterium micro-explosions, as it was for the first time achieved with the Teller-Ulam configuration for a deuterium macro-explosion in the 1952 Mike test.

**Acknowledgment**

I would like to express my sincere thanks to Dr. Stephan Fuelling for his valuable comments and suggestions, and for the real artwork he has done.

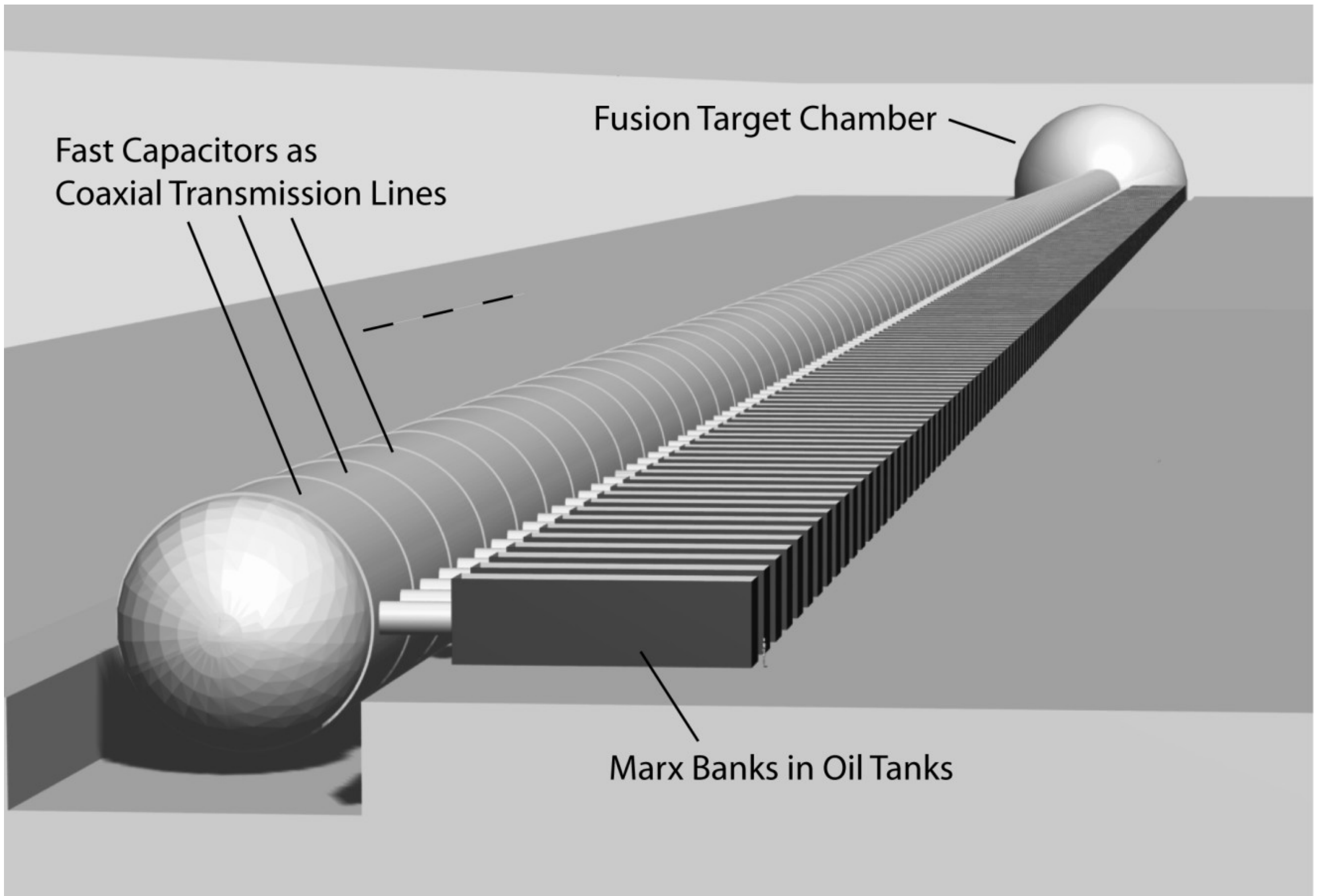

**Figure 5.** Artistic perception of a 1.5 km long Super Marx generator, composed of 100x 15 m long high voltage capacitors each designed as a magnetically insulated coaxial transmission line. The coaxial capacitors/transmission lines are placed inside a large vacuum vessel. Each capacitor/transmission line is charged by two conventional Marx generators up symmetrically to 10 MV (± 5 MV). After charge-up is completed, the Marx generators are electrically decoupled from the capacitors/transmission lines. The individual capacitors/transmission lines are subsequently connected in series via spark gap switches (i.e. the ‘Super Marx’ generator), producing a potential of 1 GV.

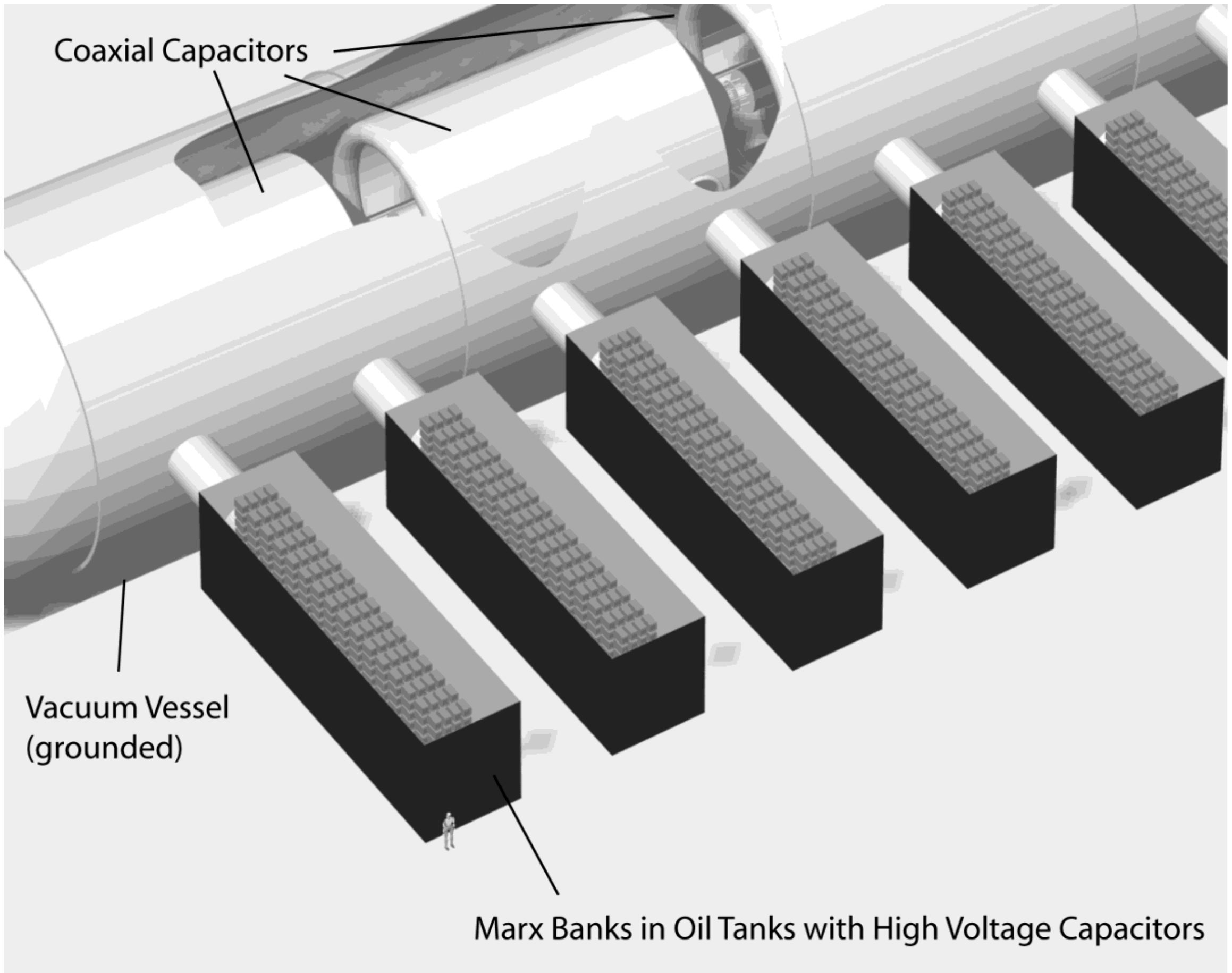

**Figure 6.** Detail view of a section of the Super Marx generator. Two conventional Marx banks charge up one coaxial capacitor/transmission line element to 10 MV.

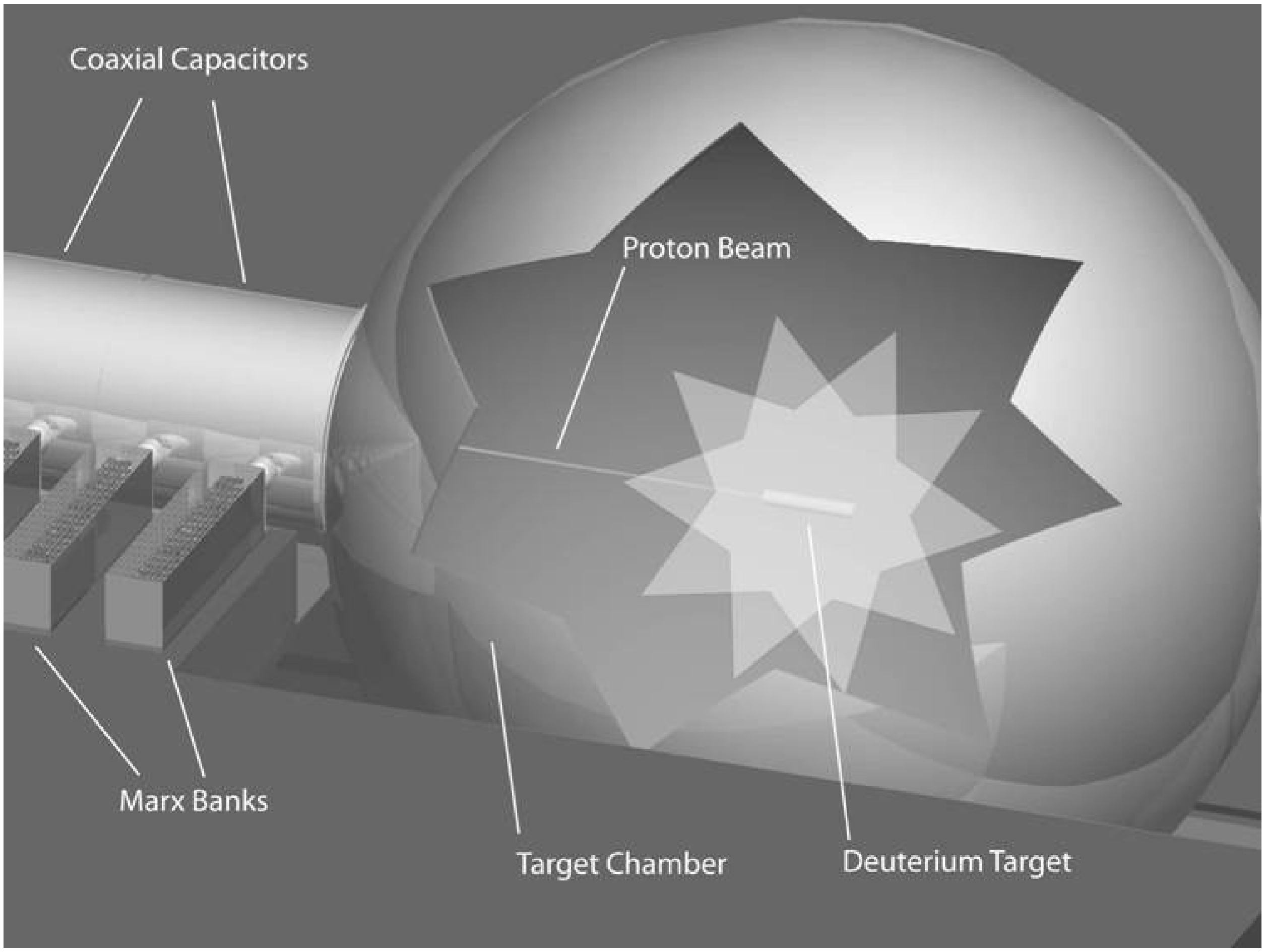

**Figure 7**. Injection of GeV – 10 MA proton beam, drawn from super Marx generator made up of magnetically insulated coaxial capacitors into chamber with cylindrical deuterium target.